\documentclass[englishUK,copyright,creativecommons]{eptcs}

\usepackage{amsmath,amssymb,wasysym,xspace}
\usepackage{mathptmx,stmaryrd}
\usepackage{eurosym}
\usepackage[T1]{fontenc}
\usepackage[mathscr]{euscript}
\usepackage[usenames,dvipsnames]{color}
\usepackage{epsfig}
\usepackage{amsfonts}
\usepackage{latexsym}
\usepackage{listings}
\usepackage{enumerate}
\usepackage{url}
\usepackage{hyperref}
\usepackage{wasysym}
\usepackage{amsthm}

\usepackage[usenames]{color}

\providecommand{\event}{Third International Workshop on \\ Behavioural Types (BEAT 2014)} 

\newcommand{\rlev}{\ensuremath{\mathsf{r}}}
\newcommand{\wlev}{\ensuremath{\mathsf{w}}}
\newcommand{\glev}{\ensuremath{\ell}}

\newdimen\proofrulebreadth \proofrulebreadth=.05em
\newdimen\proofdotseparation \proofdotseparation=1.25ex
\newdimen\proofrulebaseline \proofrulebaseline=2ex
\newcount\proofdotnumber \proofdotnumber=3
\let\then\relax
\def\hfi{\hskip0pt plus.0001fil}
\mathchardef\squigto="3A3B
\newif\ifinsideprooftree\insideprooftreefalse
\newif\ifonleftofproofrule\onleftofproofrulefalse
\newif\ifproofdots\proofdotsfalse
\newif\ifdoubleproof\doubleprooffalse
\let\wereinproofbit\relax
\newdimen\shortenproofleft
\newdimen\shortenproofright
\newdimen\proofbelowshift
\newbox\proofabove
\newbox\proofbelow
\newbox\proofrulename
\def\shiftproofbelow{\let\next\relax\afterassignment\setshiftproofbelow\dimen0 }
\def\shiftproofbelowneg{\def\next{\multiply\dimen0 by-1 }%
\afterassignment\setshiftproofbelow\dimen0 }
\def\setshiftproofbelow{\next\proofbelowshift=\dimen0 }
\def\setproofrulebreadth{\proofrulebreadth}

\def\prooftree{
\ifnum  \lastpenalty=1
\then   \unpenalty
\else   \onleftofproofrulefalse
\fi
\ifonleftofproofrule
\else   \ifinsideprooftree
        \then   \hskip.5em plus1fil
        \fi
\fi
\bgroup
\setbox\proofbelow=\hbox{}\setbox\proofrulename=\hbox{}%
\let\justifies\proofover\let\leadsto\proofoverdots\let\Justifies\proofoverdbl
\let\using\proofusing\let\[\prooftree
\ifinsideprooftree\let\]\endprooftree\fi
\proofdotsfalse\doubleprooffalse
\let\thickness\setproofrulebreadth
\let\shiftright\shiftproofbelow \let\shift\shiftproofbelow
\let\shiftleft\shiftproofbelowneg
\let\ifwasinsideprooftree\ifinsideprooftree
\insideprooftreetrue
\setbox\proofabove=\hbox\bgroup$\displaystyle 
\let\wereinproofbit\prooftree
\shortenproofleft=0pt \shortenproofright=0pt \proofbelowshift=0pt
\onleftofproofruletrue\penalty1
}

\def\eproofbit{
\ifx    \wereinproofbit\prooftree
\then   \ifcase \lastpenalty
        \then   \shortenproofright=0pt  
        \or     \unpenalty\hfil         
        \or     \unpenalty\unskip       
        \else   \shortenproofright=0pt  
        \fi
\fi
\global\dimen0=\shortenproofleft
\global\dimen1=\shortenproofright
\global\dimen2=\proofrulebreadth
\global\dimen3=\proofbelowshift
\global\dimen4=\proofdotseparation
\global\count255=\proofdotnumber
$\egroup  
\shortenproofleft=\dimen0
\shortenproofright=\dimen1
\proofrulebreadth=\dimen2
\proofbelowshift=\dimen3
\proofdotseparation=\dimen4
\proofdotnumber=\count255
}

\def\proofover{
\eproofbit 
\setbox\proofbelow=\hbox\bgroup 
\let\wereinproofbit\proofover
$\displaystyle
}%
\def\proofoverdbl{
\eproofbit 
\doubleprooftrue
\setbox\proofbelow=\hbox\bgroup 
\let\wereinproofbit\proofoverdbl
$\displaystyle
}%
\def\proofoverdots{
\eproofbit 
\proofdotstrue
\setbox\proofbelow=\hbox\bgroup 
\let\wereinproofbit\proofoverdots
$\displaystyle
}%
\def\proofusing{
\eproofbit 
\setbox\proofrulename=\hbox\bgroup 
\let\wereinproofbit\proofusing
\kern0.3em$
}

\def\endprooftree{
\eproofbit 
  \dimen5 =0pt
\dimen0=\wd\proofabove \advance\dimen0-\shortenproofleft
\advance\dimen0-\shortenproofright
\dimen1=.5\dimen0 \advance\dimen1-.5\wd\proofbelow
\dimen4=\dimen1
\advance\dimen1\proofbelowshift \advance\dimen4-\proofbelowshift
\ifdim  \dimen1<0pt
\then   \advance\shortenproofleft\dimen1
        \advance\dimen0-\dimen1
        \dimen1=0pt
        \ifdim  \shortenproofleft<0pt
        \then   \setbox\proofabove=\hbox{%
                        \kern-\shortenproofleft\unhbox\proofabove}%
                \shortenproofleft=0pt
        \fi
\fi
\ifdim  \dimen4<0pt
\then   \advance\shortenproofright\dimen4
        \advance\dimen0-\dimen4
        \dimen4=0pt
\fi
\ifdim  \shortenproofright<\wd\proofrulename
\then   \shortenproofright=\wd\proofrulename
\fi
\dimen2=\shortenproofleft \advance\dimen2 by\dimen1
\dimen3=\shortenproofright\advance\dimen3 by\dimen4
\ifproofdots
\then
        \dimen6=\shortenproofleft \advance\dimen6 .5\dimen0
        \setbox1=\vbox to\proofdotseparation{\vss\hbox{$\cdot$}\vss}%
        \setbox0=\hbox{%
                \advance\dimen6-.5\wd1
                \kern\dimen6
                $\vcenter to\proofdotnumber\proofdotseparation
                        {\leaders\box1\vfill}$%
                \unhbox\proofrulename}%
\else   \dimen6=\fontdimen22\the\textfont2 
        \dimen7=\dimen6
        \advance\dimen6by.5\proofrulebreadth
        \advance\dimen7by-.5\proofrulebreadth
        \setbox0=\hbox{%
                \kern\shortenproofleft
                \ifdoubleproof
                \then   \hbox to\dimen0{%
                        $\mathsurround0pt\mathord=\mkern-6mu%
                        \cleaders\hbox{$\mkern-2mu=\mkern-2mu$}\hfill
                        \mkern-6mu\mathord=$}%
                \else   \vrule height\dimen6 depth-\dimen7 width\dimen0
                \fi
                \unhbox\proofrulename}%
        \ht0=\dimen6 \dp0=-\dimen7
\fi
\let\doll\relax
\ifwasinsideprooftree
\then   \let\VBOX\vbox
\else   \ifmmode\else$\let\doll=$\fi
        \let\VBOX\vcenter
\fi
\VBOX   {\baselineskip\proofrulebaseline \lineskip.2ex
        \expandafter\lineskiplimit\ifproofdots0ex\else-0.6ex\fi
        \hbox   spread\dimen5   {\hfi\unhbox\proofabove\hfi}%
        \hbox{\box0}%
        \hbox   {\kern\dimen2 \box\proofbelow}}\doll%
\global\dimen2=\dimen2
\global\dimen3=\dimen3
\egroup 
\ifonleftofproofrule
\then   \shortenproofleft=\dimen2
\fi
\shortenproofright=\dimen3
\onleftofproofrulefalse
\ifinsideprooftree
\then   \hskip.5em plus 1fil \penalty2
\fi
}

\newtheorem{definition}{Definition}[section]

\newtheorem{theorem}[definition]{Theorem}
\newtheorem{corollary}[definition]{Corollary}
\newtheorem{proposition}[definition]{Proposition}

\newtheorem{example}[definition]{Example}

\newcommand{\myrule}[3]{\begin{prooftree}
 #1 \justifies   #2   \using{\rln{#3}} \end{prooftree}}
 \newcommand{\myformula}[1]{\\[5pt]\centerline{#1}\\[5pt]}
  \newcommand{\myformulaD}[1]{\\[5pt]\centerline{#1}\\[0pt]}
   \newcommand{\myformulaN}[1]{\\[5pt]\centerline{#1}\\[-8pt]
   }

  \newenvironment{myitemize}
               {\begin{itemize}\vspace{-5pt}\topsep0pt\parskip0pt\partopsep0pt\itemsep0pt\leftmargin0pt\itemsep2pt\labelwidth0pt\labelsep3pt}
               {\end{itemize}}
               
\newenvironment{mytable}
               {\begin{table}}{\vspace{-7pt}\end{table}}

  \newenvironment{mydefinition}
               {\begin{definition}}{\end{definition}}

\newcommand{\salta}[1]{}

\newcommand{\A}{{\mathcal A}}

\newcommand{\E}{{\mathcal E}}

\newcommand{\pc}{~|~}

\newcommand{\red}{\longrightarrow}

\newcommand{\set}[1]{\{#1\}}

\newcommand{\labelx}[1]{\label{#1}}

\newcommand{\wtproc}[4]{\Gamma\vdash #1 \rhd #2\dup #3}
\newcommand{\wtprocG}[5]{#1 \vdash #2 \rhd #3\dup #4}

\newcommand{\Gvti}[4]{\ensuremath{#1\to\pset:\{#2_i({#3}_i). #4_i \}_{i \in I}}}    

\newcommand{\pp}{{\ensuremath{\mathtt{p}}}}
\newcommand{\q}{{\ensuremath{\mathtt{q}}}}
\newcommand{\rp}{{\ensuremath{\mathtt{r}}}}
\newcommand{\G}{\ensuremath{{\sf G}}}
\newcommand{\Lg}{\ensuremath{{\sf L}}}
\newcommand{\ST}{\ensuremath{S}}
\newcommand{\T}{\ensuremath{\mathsf{T}}}
\newcommand{\pT}{\ensuremath{T}}
\renewcommand{\P}{\ensuremath{P}}
\newcommand{\N}{\ensuremath{N}}
\newcommand{\pset}{\ensuremath{\q}}

\newcommand{\ty}{\textbf{t}}
\newcommand{\End}{\kf{end}}
\newcommand{\next}{\kf{next}}
\newcommand{\true}{\kf{true}}
\newcommand{\false}{\kf{false}}
\newcommand{\new}{\kf{new}}
\newcommand{\inact}{\ensuremath{\mathbf{0}}}
\newcommand{\kf}[1]{\ensuremath{\mathsf{#1}\xspace}}
\newcommand{\Bool}{\kf{bool}}
\newcommand{\Nat}{\kf{nat}}
\renewcommand{\true}{\kf{true}}
\renewcommand{\false}{\kf{false}}
\newcommand{\e}{\kf{e}}
\newcommand{\x}{x}
\newcommand{\val}{v}
\newcommand{\vo}{u}
\newcommand{\eval}[2]{#1 \downarrow #2}

\newcommand{\recvar}{\ensuremath{X}}

\newcommand{\cond}[3]{\kf{if}~ #1 ~\kf{then} ~#2 ~\kf{else}~#3}
\newcommand{\sep}{\ensuremath{~\mathbf{|\!\!|}~ }}

\renewcommand{\l}{\lambda}
\newcommand{\M}{\mathcal{M}}

\newcommand{\sendMg}[2]{{#1}!\set{\lambda_i (#2_i). \M_i}_{i \in I}}
\newcommand{\recMg}[2]{{#1}?\set{\lambda_i (#2_i). \M_i}_{i \in I}}

\newcommand{\ch}{\mathsf{s}}
\newcommand{\cc}{\mathsf{c}}
\newcommand{\que}{h}

\newcommand{\stackred}[1]{\xrightarrow{#1}}
\newcommand{\proj}[2]{#1 \!  \upharpoonright  \! #2\,}
\newcommand{\rest}[1]{(\nu\,{#1})}
\newcommand{\procset}{\mathcal{P}}

\newcommand{\Q}{\ensuremath{Q}}
\newcommand{\sub}[2]{\set{#1/#2}}
\newcommand{\labin}{lin}
\newcommand{\labout}{lout}
\newcommand{\leqt}{\leq}
\newcommand{\imply}{\text{ implies }}
\newcommand{\Tset}{\mathcal{T}}

\newcommand{\cuni}{\Cup}   

\newcommand{\typout}[4]{ ! #2(#3). #4}
\newcommand{\typin}[4]{ ? #2(#3).#4}
\newcommand{\typelout}[2]{  ! #2}
\newcommand{\typelin}[2]{ ? #2}
\newcommand{\procin}[4]{#1 ?  #3 .#4}
\newcommand{\procout}[4]{#1 ! #3. #4}
\newcommand{\agree}{\propto}
\newcommand{\mt}[1]{|#1|}

\newcommand{\h}{\ensuremath{h}}

\newcommand{\pt}{\kf{part}}

\newcommand{\rln}[1]{\textsc{#1}}
\newcommand{\dup}{ :}

\newcommand{\eq}{\text{\o}}
\newcommand{\mess}[3]{(#1, #2,#3)}

\DeclareMathAlphabet{\mathpzc}{OT1}{pzc}{m}{it}

\newcommand{\nonce}[1]{\mathpzc{nonce}_{#1}}

\definecolor{ocre}{rgb}{.92,.86,.3}%
\definecolor{verde}{rgb}{0.50,0.70,0.4}
\definecolor{beppeblue}{rgb}{0,0,0.4}
\definecolor{lucagreen}{rgb}{0,0.3,0}
\definecolor{lightgreen}{rgb}{0.79,0.86,.79}
\definecolor{lesslightgreen}{rgb}{0.39,.9,.39}
\definecolor{mydarkgreen}{rgb}{0.10,0.43,0}
\definecolor{darkred}{rgb}{0.67,0.27,0.39}
\definecolor{ultralightgray}{gray}{0.85}
\definecolor{midgray}{gray}{0.6}
\definecolor{siena}{rgb}{0.58,0.23,0.34}
\definecolor{mgreen}{rgb}{0,0.5,0}
\definecolor{lucared}{rgb}{0.8,0,0}

\newcommand{\pskip}[1]{}

\newcommand{\Nonces}{\mathit{Nonces}}

\title{Self-Adaptation and Secure Information Flow in Multiparty Structured Communications: A Unified Perspective
}
\author{Ilaria Castellani
\institute{INRIA Sophia-Antipolis (FR)}
\and
Mariangiola Dezani-Ciancaglini
\institute{Universit\`{a} di Torino (IT)}
\and
Jorge A. P\'{e}rez
\institute{University of Groningen (NL)}
}

\def\titlerunning{Self-Adaptation and Secure Information Flow in Multiparty Structured Communications}
\def\authorrunning{I. Castellani \& M. Dezani-Ciancaglini \& J. A. P\'{e}rez}

\begin{document}
\maketitle

\begin{abstract}
  We present initial results on a comprehensive model of
  structured communications, in which 
  self-adaptation and security concerns are jointly addressed. More
  specifically, we propose a model of self-adaptive, multiparty
  communications with secure information flow
  guarantees. 
 In this model, 
security violations occur when processes attempt to
  read or write messages of inappropriate security levels within directed
  exchanges. Such violations 
trigger adaptation mechanisms that
  prevent the violations to occur and/or to propagate their effect in
  the choreography.  Our
  model 
  is equipped with local and global mechanisms for reacting to
  security violations; type soundness results ensure that global
  protocols are still correctly executed, while the system adapts
    itself to preserve security.
\end{abstract}

\section{Introduction}
Large-scale distributed systems are nowadays conceived as heterogeneous collections of software artifacts. Hence, \emph{communication} plays a central role in their overall behavior. In fact, ensuring that the different components follow the stipulated protocols is a basic requirement in certifying system correctness. However, as communication-centric systems arise in different computing contexts,  system correctness can no longer be characterized solely in terms of protocol conformance. Several other aspects ---for instance, security,  evolvability/adaptation, explicit distribution, time--- are becoming increasingly relevant in the specification of actual interacting systems, and should be integrated into their correctness analysis. Recent proposals have addressed some of these aspects, thus extending the applicability of known reasoning techniques over models of communication-based systems. In the light of such proposals, a pressing challenge consists in understanding whether known models and techniques, often devised in isolation, can be harmoniously integrated into unified frameworks.

As an  example, 
consider the multiparty interaction between a user, his bank, a store, and a social network.  All exchanges occur on top of a browser, which relies on plug-ins to integrate information from different services.  For instance, a plug-in may announce in the social network that the user has just bought an item from the store.  That is, agreed exchanges between the user, the bank, and the store may in some cases lead to a (public) message announcing the transaction. We would like to ensure that the buying protocol works as expected, but also to avoid that sensitive information, exchanged in certain parts of the protocol, is leaked ---e.g., in a \emph{tweet} which mentions the credit card used in the transaction.  Such an undesired behavior should be corrected as soon as possible. In fact, we would like to stop
relying on the (unreliable) participant in ongoing/future instances of the protocol. Depending on how serious the leak is, however, we may also like to react in different ways. If the leak is minor (e.g., because the user interacted incorrectly with the browser), then we may simply identify the source of the leak and postpone the reaction to a later stage,
enabling unrelated participants in the choreography to proceed with their exchanges. Otherwise, if the leak is serious (e.g., when the plug-in is compromised by a malicious participant) we 
may wish to adapt the choreography as soon as possible, removing the plug-in and modifying the behavior of the involved participants.
This form of reconfiguration, however, should only concern the
participants involved with the insecure plug-in; 
participants not directly affected by the leak should not be unnecessarily restarted. In our example, since the unintended tweet concerns only the user, the store and the social network, the update should not affect the behavior of the bank.

To analyze such choreographic scenarios, we propose a
framework for self-adaptive, multiparty communications which ensures
basic guarantees for access control and secure information
flow.
%
The framework consists of 
a language for 
processes and networks, 
global types, and runtime monitors.
Runtime monitors are obtained as projections from global types onto individual participants.
Processes represent code that will be coupled with monitors to implement participants.
A network is a collection of monitored processes which realize a choreography  as described by the global type.

%

Intuitively, a monitor defines the behavior of a single participant in
the choreography. In our proposal, the monitor also defines a security
policy by stipulating \emph{reading and writing permissions},
represented by \emph{security levels}. The reading permission is an
upper bound for the level of incoming messages, and the writing
permission is a lower bound for the level of outgoing messages. A
reading or writing violation occurs when a participant attempts to
read or write a message whose level is not allowed by the
corresponding reading or  writing permission.
A monitored operational semantics for networks is given by a reduction relation which ensures that the reading/writing permissions are respected or, in case they are violated, that an appropriate adaptation mechanism is triggered  to limit the impact of the violation. 

We consider both \emph{local} and \emph{global}
 adaptation mechanisms,  intended to handle minor and serious leaks,
 respectively.  The local mechanism works as follows: in case of
 a reading violation, the behavior of the monitor is modified so as to
 omit the disallowed read, and a process compliant with the new
 monitor is injected; in case of a writing violation, we
   penalize the sender by decreasing the reading level of his monitor and the implementation for the receiver is replaced.
   (In any case, the culprit of a reading/writing violation is always considered to be the
   sender.\footnote{This is because the sender has an {\em active role}
   in producing and disseminating information through the system,
   while the receiver only has a passive role, and thus cannot be
   blamed for finding a sensitive value in the queue.})
 The global
 mechanism relies on distinguished low-level values called
 \emph{nonces}. 
When 
an attempt to leak a 
value is detected, the value is replaced in the communication with a
fresh
nonce. This avoids improperly communicating the
value and allows the whole system to make progress, for
 the benefit of the participants not involved in the violation. The
 semantics may then trigger at any point a reconfiguration action 
 which removes the whole group of participants 
that may propagate the nonce
and replaces it with a new choreography (global type).
 Thus,  in this form of adaptation, one part of the choreography is isolated
 and replaced.
 


\section{Syntax}\label{s:syntax}

Our calculus is inspired by that of~\cite{CDV14}, where 
security issues were not addressed
and adaptation was determined by changes of a
global state, which is not needed for our present purposes. 
We consider networks with three active components: \emph{global
  types}, \emph{monitors}, and \emph{processes}. A global type
represents the overall communication choreography 
over a set of
participants~\cite{CHY12}. Moreover, the global type defines reading
permissions for each participant, following~\cite{CCD12}. By projecting the
global type onto participants, we obtain monitors: in essence, these are
local types that define the communication protocols of the
participants. The association of a process with a ``fitting'' monitor,
dubbed {\em monitored process}, incarnates a participant whose 
process implements the monitoring protocol.  Notably, we exploit
intersection types, union types and subtyping to make this ``fitting'' 
relation more flexible.

As usual, we consider a finite lattice of {\em security levels} \cite{Denning76},
ranged over by $\glev,\glev',\ldots$. We denote by $\sqcup$ and $\sqcap$ the join
and meet operations on the lattice, and by $\bot$ and $\top$ its
bottom and top elements.
Also, we use $\rlev,\rlev',\ldots$ and $\wlev,\wlev',\ldots$ to range over 
levels denoting reading and writing permissions, respectively.

\bigskip

\noindent {\bf \emph{Global Types and Monitors.}}
Global types define overall schemes of labeled communication
between session participants.  In our setting, they also
prescribe the reading levels of the 
participants.
 We assume base sets of \emph{participants}, ranged over by $\pp, \pset, \rp, \ldots$;
\emph{labels}, ranged over by $\lambda, \lambda', \ldots$; and
\emph{recursion variables}, ranged over by $\ty, \ty', \ldots$.
We also assume a set of
basic \emph{sorts} ($\Bool, \Nat, \ldots$), ranged over by $S$.
\begin{mydefinition}[Global Types and Security Global Types]\labelx{gtypesdef}

{\em Global types} are defined by:
\myformulaN{
\begin{tabular}{rclrrrclr}
 $\G$ & $::=$ & $\Gvti \pp \lambda \ST {\G}$  & \sep & $\ty$ & \sep & $\mu\ty. \G$ &\sep &$\End$
\end{tabular}}
\smallskip

\noindent
We let $\pt(\G)$ denote the set of participants in $\G$,
i.e., all senders $\pp$ 
and receivers $\q$ 
occurring in
$\G$. 
A {\em security global type} is a pair
  $(\G,\Lg)$, where $\G$ is a global type and $\Lg$ maps each 
  $\pp$
  in $\pt(\G)$
to a reading level \rlev.
\end{mydefinition} 
A global type describes a sequence of value exchanges. Each value
  exchange is directed between a sender $\pp$ and a receiver $\pset$,
  and characterized by a label $\l$, which represents a choice among
  different alternatives.  In writing $\Gvti \pp \lambda \ST {\G}$ we
  implicitly assume that $\pp \neq \pset$ and $\l_i \neq \l_j$
  for all $i\neq j$.
%
%
The global type $\End$ denotes the completed choreography.
To account for recursive protocols, we consider recursive global types. As customary, 
we require guarded recursions and we adopt an equi-recursive view of recursion for all syntactic
categories,
identifying a recursive definition with its unfolding.

Monitors are obtained as \emph{projections} from global types onto
individual participants, following standard
definitions~\cite{CHY08,BCDDDY08}.  
The projection of a global
  type $\G$ onto participant $\pp$, denoted $\proj{\G}\pp$, generates
  the monitor for $\pp$. As usual, in order for $\proj{\G}\pp$ to be
  defined, it is required that whenever $\pp$ is not involved in some
  directed communication of $\G$, it has equal projections in the
  different branchings of that communication.  We say  $\G$ is
  {\em well formed} if the projection $\proj{\G}\pp$ is defined
  for all $\pp \in \pt(\G)$. 
  In the following we assume that all
(security) global types are well formed.

Although monitors can be seen as local types, in our
model 
they have an active role in
the dynamics of networks, since they guide and enable/disable directed communications.
\begin{mydefinition}[Monitors]\labelx{monitorsdef} The set of \emph{monitors} is defined by:
\myformulaN{
\begin{tabular}{rcllllllllr}
 $\M $ & $::=$ & $\recMg{\pp}{S}$  & $\sep$ &  $\sendMg{\pset}{S}$& \sep & $\ty$ & \sep & $\mu\ty. \M$& $\sep$   &$\End$
\end{tabular}
}
\end{mydefinition}

An input monitor $\recMg{\pp}{S}$ fits 
with a process that can
receive, for each $i\in I$, a value of sort $\ST_i$, labeled by
$\lambda_i$, and then continues as specified by $\M_i$.
This corresponds to an external choice. 
Dually, an output monitor $\sendMg{\pset}{S}$ fits with a process
which can send, for each $i\in I$, a value of sort $\ST_i$,
labeled by $\lambda_i$, 
and then continues as prescribed by $\M_i$. As such, it corresponds to an internal choice. 


\bigskip

\noindent {\bf \emph{Processes and Networks.}}
We assume a set of \emph{expressions}, ranged over by $\e, \e',
\ldots$, which includes booleans and naturals (with operations over
them) and a denumerable set $Nonces = \{\nonce{i} \sep  i\geq0\}$. 
An expression $\nonce{i}$ ---where $i$ is fresh--- is a dummy value that
is generated at
runtime to be used in place of some improperly sent value, in order to prevent security violations, see~\S\,\ref{semanticSection}.
Each expression $\e$ is equipped with a security level, denoted
$lev(\e)$; 
for every $i$, $lev(\nonce{i}) = \bot$. 
Values are ranged over by $\val,\val'$; we use $\vo$ to denote an {\em
  extended value}, which is either a value or a nonce.

We now define our set of processes, which represent code that will be coupled with monitors to implement participants.
Our model, like~\cite{CDV14}--- but unlike other session
calculi~\cite{HKV98,CHY08,BCDHY13,CDPY13}--- uses processes that do not specify their partners in communication actions. 
It is the associated monitor which determines the partner in a given communication.
Thus, processes represent flexible code that can be associated with
different monitors to incarnate different  
participants. 
Communication actions 
are performed through \emph{channels}.  Each process owns a unique
channel, which by convention is denoted by $y$ in the user
  code.
At runtime,
channel $y$ will be replaced by a {\em session channel} $\ch[\pp]$, where
$\ch$ is the session name and $\pp$ denotes the participant. We
  use $\cc$ to stand for a user channel $y$ or a session channel
$\ch[\pp]$. 



\begin{mydefinition}[Processes]\labelx{procdefinition}
The set of \emph{processes} is defined by:
\myformulaN{
\begin{tabular}{rclll}
 $\P $ & $::=$ & $  \inact  ~~~\sep~~~  \procin  \cc  \pp {\lambda(\x)} \P  ~~~\sep~~~  \procout \cc \pset {\lambda(\e)} \P ~~~\sep~~~ \recvar ~~~\sep~~~   \mu\recvar.\P $  ~~~\sep~~~  $\cond{\e}{\P}{\P} ~~ \sep ~~ \P +\P$  
 \end{tabular}
}
\end{mydefinition}
The syntax of processes is rather standard: in addition to usual
constructs for communication, recursion and conditionals, it includes
the operator $+$, which represents external choice.  For instance,
$\procout \cc \pset \lambda(\e) \P$ denotes a process which sends along
$\cc$ label $\lambda$ and the value of the expression $\e$ and then
behaves like $\P$.
We assume the following precedence among operators: prefix, external choice, recursion.

The previously introduced entities (global types, monitors, processes)
are used to define \emph{networks}. 
A network is a
collection of monitored processes which realize a choreography as
described by a global type.  The choreography is initiated by the
``$\new$'' construct applied to a security global type
$(\G,\Lg)$. 
This construct, {akin to} a {\em session initiator}~\cite{CDV14}, 
is denoted $\new(\G,\Lg)$. 
  In carrying on a multiparty interaction, a
process is always controlled by a monitor, which ensures that all
{its communications}
agree with 
the protocol prescribed by the global type. 
Each
monitor 
is equipped with a reading permission $\rlev$ and a writing permission
$\wlev$.
A monitored process, written $\M^{\rlev,\wlev}[\P]$, denotes a process \P\
controlled by a monitor $\M$. 

Data are exchanged among participants asynchronously, by
means of {\em message queues}, ranged over by $h, h', \ldots$.  There
is one such queue for each active session. We denote by $\ch\dup\h$
the {\em named queue} associated with session $\ch$.
 The empty queue
is denoted by $\eq$.  Messages in queues are of the form $\mess\pp
\pset {\l(\vo)}$, indicating that the label $\l$ and the extended value 
$\vo$ are communicated with sender $\pp$ and receiver
$\pset$. Queue concatenation is denoted by
``$\cdot$'':  it is associative and has $\eq$ as neutral element.  

The parallel composition of session initiators, monitored processes, and runtime queues forms a network. 
Networks can be restricted on session names.

\begin{mydefinition}[Networks]\labelx{netdef}
The set of {\em networks} is defined by:
\myformulaN{ $\N ~ ::= ~\new(\G,\Lg) ~\sep~\M^{\rlev,\wlev}[\P]~\sep~ \ch\dup\h ~\sep~ \N\pc\N  ~\sep~ (\nu\ch)\N$}
\end{mydefinition}
As mentioned above, 
annotations $\rlev$ and $\wlev$
in 
$\M^{\rlev,\wlev}[\P]$
 represent reading and writing
permissions for process $P$.  While $\rlev$ acts as an upper bound
for reading, $\wlev$ acts as a lower bound for writing.  When the
choreography is initialized, the reading level is set according to map
$\Lg$; the writing level is always set to~$\bot$.  
The actions performed by the process determine dynamic modifications to 
these levels.
In writing monitored processes we omit the levels when they
are not used. Also, we shall sometimes write
  $\M^{\rlev,\wlev}_\pp[\P]$ (or simply $\M_\pp[\P]$) to indicate that
  the channel in $\P$ is $\ch[\pp]$ for some $\ch$.


\medskip

As in~\cite{CDV14}, process types (called   \emph{types} when not ambiguous)
describe process communication behaviors.  Types have
prefixes corresponding to input and output actions. In particular, an
{\em input type} (resp. {\em output type}) is a type whose 
prefix corresponds to an input (resp. output) action,
while the {\em continuation} of a type is the type following its first 
prefix. A {\em communication type} is either an input or an output
type.  Intersection types are used to type external choices, since an
external choice offers both behaviors of the composing
processes. Dually, union types are used to type conditional
expressions (internal choices).

To formally define types, we first give the more liberal syntax of
\emph{pre-types} and then we characterize process types by fixing some
natural restrictions on pre-types.

\begin{mytable}
  \normalsize
\[\begin{array}{ccc}
\multicolumn{3}{c}{ \labin(\typin \pp \lambda \ST \pT)  = \labout(\typout \pset \lambda \ST \pT)  =  \set{\lambda}}
 \\[2pt]
\multicolumn{3}{c}{\labin(\typout \pset \lambda \ST \pT)  = \labin(\typelout \pset \l)  =   \labout(\typin \pp \lambda \ST \pT)  = \labout(\typelin{\pset}{\l})  =  \emptyset}\\[2pt]
\multicolumn{3}{c}{\labin(\pT_1 \wedge \pT_2)  = \labin(\pT_1 \vee \pT_2)  =   \labin(\pT_1) \cup \labin(\pT_2) }\\[2pt]
\multicolumn{3}{c}{\labout(\pT_1 \wedge \pT_2)  =  \labout(\pT_1 \vee \pT_2)  =  \labout(\pT_1) \cup \labout(\pT_2)}
\end{array}\]
\caption{ \normalsize The mappings  $\labin$ and $\labout$, as required in Definition~\ref{tc}.}\labelx{linlout}
\end{mytable}

\begin{mydefinition}[Pre-types]
The set of \emph{pre-types} is inductively defined by:
\myformula{$\pT ~  ::=  ~ \typin{\pp}\lambda \ST \pT   \sep
      \typout \pset \lambda \ST \pT  
       \sep\pT \wedge \pT \sep
        \pT \vee \pT  \sep \ty\sep\mu\ty.\pT\sep\End$}
where $\wedge$ and $\vee$ are considered modulo idempotence, commutativity, and associativity.
\end{mydefinition}
\noindent
In writing pre-types and types we assume that `$.$' has precedence over `$\wedge$' and `$\vee$'.

In order to define types for processes, we have to avoid intersection
between input types with the same first label, which would represent
an ambiguous external choice: indeed, the types following a same input
prefix could be different and this would lead to a communication
mismatch. For the same reason, process types cannot contain
intersections between output types with the same label. Since we have
to match types with monitors, where internal choices are always taken
by participants sending a label, we force unions to take as arguments
output types (possibly combined by intersections or unions).
Therefore, we formalize the above restrictions by means of two
mappings from pre-types to sets of labels (Table \ref{linlout}) and
then we define types by using these mappings.

\begin{mydefinition}[Process Type]\labelx{tc}
  A \emph{(process) type} is a pre-type satisfying the following
  constraints modulo idempotence, commutativity and associativity of
  unions and intersections:

\begin{myitemize}
\item all occurrences of the shape $\pT_1 \wedge \pT_2$ are such that $\,\labin(\pT_1)\cap \labin(\pT_2) = \labout(\pT_1)\cap \labout(\pT_2) = \emptyset$.
    \item all occurrences of the shape $\pT_1 \vee \pT_2$ are such
      that $\,\labin(\pT_1) = \labin(\pT_2) = \labout(\pT_1) \cap \labout(\pT_2) = \emptyset$.
\end{myitemize}
We use $\T$ to range over types and $\Tset$ to denote the set of types. 
\end{mydefinition}
\noindent
For instance,  $(\pT\wedge\pT)\vee\pT$ is a type, whenever $\pT$ is a type, since types are considered modulo  idempotence.

We now introduce the type system for processes.
 An \emph{environment} $\Gamma$ is a finite mapping from expression variables to sorts and from process variables to types:
 \myformula{$\Gamma::=\emptyset~\sep~\Gamma,\x\dup S~\sep~\Gamma,\recvar \dup \T$} where the notation $\Gamma, x\dup S$ (resp. $\Gamma, \recvar\dup \T$) means that $\x$ (resp. $X$) does not occur in $\Gamma$.

\begin{table} [!t]
\[\begin{array}{c}
\wtproc{\inact}\cc{\End}{\Delta} ~~\rln{end}\qquad
 \wtprocG{\Gamma,\recvar:\T} {\recvar} {\cc}{\T}{\set{(\l, \recvar)}} ~~\rln{rv} \\ \\
\myrule{\wtprocG{\Gamma,\recvar:\T} \P \cc \T \Delta}
 {\wtproc {\mu\recvar. \P}{\cc}{\T}{\Delta\sub{\T}{\recvar}}}{rec}\qquad
  \myrule {\wtprocG{\Gamma,x:\ST}{\P}{\cc}{\T}{\Delta}}
{\wtproc{\procin \cc \pp {\lambda(\x)} \P}{\cc}{\typin \pp \lambda \ST  \T} {\Delta}}{rcv}\qquad
 \myrule{\wtproc{\P} \cc {\T} {\Delta}~~~~~~~~~\Gamma\vdash \e:\ST}
{\wtproc{\procout \cc \pset{\lambda(\e)} \P}{\cc}{\typout\pset \lambda \ST \T}{\Delta}}{send}
\\ \\
 \myrule{\Gamma \vdash \e : \Bool\quad\wtproc{\P_1}{\cc}{\T_1}{\Delta_1}\quad\wtproc{\P_2}{\cc}{\T_2}{\Delta_2}\quad\T_1 \vee \T_2\in \Tset}
 {\wtproc{\cond \e {\P_1}{\P_2}}{\cc}{\T_1 \vee \T_2}{\Delta_1 \cuni \Delta_2}}{if} \\\\
 \myrule{\wtproc{\P_1}{\cc}{\T_1}{\Delta_1}\quad\wtproc{\P_2}{\cc}{\T_2}{\Delta_2}\quad\T_1 \wedge \T_2\in \Tset}
 {\wtproc{\P_1 + \P_2}{\cc}{\T_1 \wedge \T_2}{\Delta_1 \cuni \Delta_2}}{choice}
 \end{array}\]
\caption{Typing Rules for Processes.}\labelx{tr}
\end{table}

Typing rules for processes are given in Table \ref{tr}. We assume that
expressions are typed by sorts, as usual, and a nonce has all
sorts. 
In rules \rln{if} and \rln{choice} we require that the applications of
union and intersection on two types form a type (cf. conditions $\T_1
\vee \T_2\in \Tset$ and $\T_1 \wedge \T_2\in \Tset$).

The compliance between process types and monitors (\emph{adequacy}) is
made flexible by using the {\em subtyping} relation on types, denoted
$\leqt$ and defined in Table \ref{tam}. Subtyping is monotone, for
input/output prefixes, with respect to continuations and it follows
the usual set theoretic inclusion of intersection and union. Notice
that we use a weaker definition than standard subtyping on
intersection and union types, since it is sufficient to define
subtyping on types.  Intuitively,
 $\T_1 \leqt \T_2$  means that 
a process with type $\T_1$ has all the behaviors required by type $\T_2$ but possibly more.

An input monitor naturally corresponds to an external choice, while an
output monitor naturally corresponds to an internal choice. Thus,
intersections of input types are adequate for input monitors and
unions of output types are adequate for output monitors. Formally, 
\emph{adequacy} is defined as follows:


\begin{mytable}[t!]
 \normalsize
We define $\leqt$ as the minimal reflexive and transitive  relation on $\Tset$  such that:
\myformula{$
\begin{array} {cccc}
\multicolumn{3}{c}{\ty\leqt\ty\qquad \T\leqt\End\qquad\T_1 \wedge \T_2 ~\leqt~ \T_i\qquad
\T_i \leqt \T_1 \vee \T_2~~~~(i=1,2)}\\
\multicolumn{3}{c}{\T_1 \leqt \T_2 \imply  \typout \pset \lambda \ST {\T_1} \leqt
\typout \pset \lambda \ST {\T_2} \text{  and  } \typin \pp \lambda \ST {\T_1} \leqt \typin \pp \lambda \ST {\T_2}}\\
\multicolumn{3}{c}{\T \leqt \T_1\text{ and }  \T\leqt \T_2~\text{imply} ~ \T ~\leqt~ \T_1 \wedge \T_2} \\
\multicolumn{3}{c}{\T_1 \leqt \T\text{ and }  \T_2 \leqt \T~\text{imply} ~ \T_1 \vee \T_2\leqt \T}\\
\multicolumn{3}{c} {(\T_1 \vee \T_2)\wedge \T_3 \leqt \T ~\text{ iff } ~ \T_1\wedge \T_3 \leqt \T ~\text{and}~ \T_2\wedge \T_3 \leqt \T  } \\
\multicolumn{3}{c} {\T  \leqt  (\T_1 \wedge \T_2)\vee \T_3 ~\text{ iff } ~  \T  \leqt  \T_1\vee \T_3~\text{and}~ \T  \leqt   \T_2\vee \T_3}\\
\multicolumn{3}{c} { \mu\ty.\T\leqt\mu\ty.\T' ~\text{ iff } ~  \T  \leqt  \T'}
 \end{array} $}
\caption{\normalsize Subtyping on Process Types.}\labelx{tam}
\end{mytable}

\begin{mydefinition}[Adequacy]\labelx{adeA}
Let the mapping $\mt{\cdot}$ from monitors to types be defined as
\myformulaD{$\begin{array}{c}
\mt{\recMg{\pp}{S}}=\bigwedge_{i \in I} \typin {\pp} {\lambda_i} {\ST_i}{\mt{\M_i}}\qquad
\mt{\sendMg{\pset}{\ST}}=\bigvee_{i \in I} \typout {\pset} {\lambda_i} {\ST_i}{\mt{\M_i}}\\[.6em]
\mt\ty=\ty \qquad \mt{\mu\ty.\M}=\mu\ty.\mt \M \qquad \mt\End=\End
\end{array}$}
We say that type $\T$  is {\em adequate} for a monitor $\M$, notation $\T\agree\M$, if  $\T\leqt \mt\M$.
\end{mydefinition}

\section{Semantics}\labelx{semanticSection}
The semantics of monitors
and processes is given by labeled transition systems (LTS), while that
of networks is given in the style of a reduction semantics.

A monitor guides the communications of a process by choosing 
its partners 
in labeled exchanges, and by allowing only some
actions among those offered by the process.

The LTS for monitors uses labels $\pp ? \lambda$ and $\pp !
\lambda$, 
and
formalizes the expected intuitions:
\myformulaD{$\begin{array}{c}
\pp ? \set{\lambda_i (\ST_i).\M_i}_{i \in I} \stackred{\pp ? \lambda_j} \M_j ~~~~~~~~~\sendMg{\pset}{S} \stackred{{\pset ! \lambda_j}} \M_j ~~~~
j\in I 
\end{array}$}

The LTS for processes, given in Table~\ref{ltsp}, is also fairly simple. 
It relies on labels 
$\ch[\pp] ? \lambda (\vo)$ (input), $\ch[\pp] ! \lambda(\vo)$
 (output), 
 and $\glev$ (security levels
 for expressions).
The labels $\ch[\pp] ? \lambda (\vo)$ and $\ch[\pp] !
    \lambda(\vo)$ are ranged over by $\alpha, \beta$.
We use
$\eval{\e}{\vo}$ to indicate that expression $\e$ evaluates to the
extended value $\vo$,
assuming $\eval{\nonce i}{\nonce i}$. When reducing a
conditional we record the level of the tested expression in
order to track information flow.
The rules for sum specify that 
choices are performed by the communication actions, while internal
computations are transparent.

\begin{mytable}
 \normalsize
\centerline{$\begin{array}{ccc}
{
\procin {\ch[\pp]} \q {\lambda(\x)} \P \stackred{\ch[\pp] ? \lambda (\vo)} \P \sub{\vo}{\x}} & \qquad & \procout{\ch[\pp]} \pset {\lambda(\e)} \P \stackred{\ch[\pp] ! \lambda(\vo)} \P \quad \eval{\e}{\vo}\\[.4em]
$\cond{\e}{\P}{\Q}$  \stackred{lev(\e)} \P ~~~~\eval{\e}{\true}  
&\qquad&
$\cond{\e}{\P}{\Q}$  \stackred{lev(\e)} \Q ~~~~\eval{\e}{\false} \\[.5em]
\P \stackred{\alpha} \P' ~\Rightarrow~ \P + \Q \stackred{\alpha} \P'&\qquad& \P \stackred{\ell} \P' ~\Rightarrow~ \P +\Q \stackred{\ell} \P'+\Q
\end{array}$}
\caption{\normalsize LTS of processes. Symmetric rules are omitted.}\labelx{ltsp}
\end{mytable}


\begin{mytable}
\[\begin{array}{c}
\myrule{\M_\pp = \proj \G \pp \qquad \forall \pp \in \pt(\G).~ (\P_\pp,\T_\pp) \in \procset ~\&~ \T_\pp\agree\M _ \pp}
{\new(\G,\Lg)  ~ \red{} \rest{\ch}~ \prod_{\pp \,\in\, \pt(\G)} (\M_\pp^{\Lg(\pp),\bot} [\P_\pp\sub{\ch[\pp]}{y}]\pc \ch:\eq) }{Init} \\[2em]
\myrule{\P \stackred{\glev}  \P'}{\M^{\rlev,\wlev}[\P]\red\M^{\rlev\,,\,\wlev\sqcup\glev}[\P']}{UpLev} 
\qquad 
 \myrule{\M_\pp \stackred{\q ? \lambda } \widehat\M_\pp \qquad
\P \stackred{\ch[\pp] ? \lambda (\vo)} \P' \qquad lev(\vo) \leq \rlev}
{\M_\pp^{\rlev,\wlev}[\P] \pc \ch:\mess\q\pp{\lambda(\vo)}\cdot\que \red{}   \widehat\M_\pp^{\rlev,\wlev}[\P'] \pc \ch:\que}{In}\\ [2.3em]
 \myrule{{ \M_\pp \stackred{{\pset ! \lambda}} \widehat\M_\pp ~~~~
\P \stackred{{\ch[\pp] ! \lambda(\vo)}}  \P'~~~~~ 
\vo \in \Nonces \text{ or } (\vo=\val \text{ and }\wlev\leq lev(\val))
}} {\M_\pp^{\rlev,\wlev}[\P] \pc \ch:\que \red{} \widehat\M_\pp^{\rlev,\wlev}[\P'] \pc \ch:\que\cdot\mess\pp \pset {\lambda(\vo)}}
{Out} \\ [2.3em]
\myrule{{ \M_\pp \stackred{{\q? \lambda}} \widehat\M_\pp ~~~~ \nonce{i} = \next(\Nonces)~~~~
\P \stackred{{\ch[\pp] ? \lambda(\nonce i)}}  \P'~~~~~ lev(\val)\not\leq\rlev 
}} {\M_\pp^{\rlev,\wlev}[\P] \pc \ch:\mess\q\pp{\lambda(\val)}\cdot\que \red{} {\widehat\M}_\pp^{\rlev,\wlev}[\P'] \pc \ch:\que}
{InGlob} \\ [2.3em]
%
\myrule{{ \M_\pp \stackred{{\pset ! \lambda}} \widehat\M_\pp ~~~~
\P \stackred{{\ch[\pp] ! \lambda(\val)}}  \P'
~~~~~ \nonce{i} = \next(\Nonces)
~~~~~ \wlev\not\leq lev(\val) 
}} {\M_\pp^{\rlev,\wlev}[\P] \pc\M_\q^{\rlev',\wlev'}[\Q]\pc \ch:\que \red{} \widehat\M_\pp^{\rlev\sqcap \rlev',\wlev}[\P'] \pc\M_\q^{\rlev',\wlev'}[\Q]\pc \ch:\que\cdot\mess\pp \pset {\lambda(\nonce{i})}}
{OutGlob} \\ [2.3em]
\myrule{\A(\set{\P_\pp\mid \pp\in \Pi}, \nonce i)=\Pi'\quad
  F(\set{\P_\pp\mid \pp\in \Pi'})=(\G,\Lg)}{\rest{\ch}~ (\prod_{\pp\in
    \Pi}\M_\pp[\P_\pp]\pc\ch:\que)\red{}\rest{\ch}~ (\prod_{\pp\in
    \Pi-\Pi'}\M_\pp[\P_\pp]\pc\ch:\que\setminus \Pi')\pc \new(\G,\Lg)
}
{Refresh}\\ [2em]
\myrule{{ \M_\pp \stackred{{\q? \lambda}}\widehat\M_\pp 
~~~~~(\P',\T) \in \procset ~~~~~ \T\agree\widehat\M_\pp
~~~~~lev(\val)\not\leq \rlev
}} {\M_\pp^{\rlev,\wlev}[\P] \pc\ch:\mess\q\pp{\lambda(\val)}\cdot\que \red{} \widehat\M_\pp^{\rlev,\wlev}[\P'] \pc \ch:\que}
{InLoc} \\ [2.3em]
\myrule{{ \M_\pp \stackred{{\pset ! \lambda}} \widehat\M_\pp ~~~~
\P \stackred{{\ch[\pp] ! \lambda(\val)}}  \P'
~~~~~\widehat\M_\q=\M_\q\setminus?(\pp,\lambda) 
~~~~~(\Q',\T) \in \procset 
~~~~~ \T\agree\widehat\M_\q
~~~~~ \wlev\not\leq lev(\val) 
}} {\M_\pp^{\rlev,\wlev}[\P] \pc\M_\q^{\rlev',\wlev'}[\Q] \red{}\widehat\M_\pp^{\,\rlev\sqcap\rlev',\wlev}[\P'] \pc \widehat\M_\q^{\rlev',\wlev'}[\Q'\sub{\ch[\q]}{y}]}
{OutLoc} 
\\ [2.3em]
 \myrule{\N_1\equiv \N_1' \quad \N_1'\red \N_2' \quad \N_2\equiv \N_2'}{\N_1\red \N_2}{Equiv}\qquad\qquad \myrule{\N\red{}\N'}{\E[\N]\red\E[\N']}{Ctx}
\end{array}\]
\caption{Reduction Rules for Networks.}\labelx{nr}
\end{mytable}

The reduction of networks assumes a collection $\procset$
of pairs $(\P,\T)$ 
of processes together with their types.  It uses
a  rather natural structural equivalence $\equiv$ 
which erases monitored processes with $\End$ monitor and commutes 
independent messages (with different senders or different
receivers) in queues~\cite{CCD12}.
The reduction rules for networks are given in Table~\ref{nr}. We briefly describe them:
\begin{enumerate}[1.]
\item Rule \rln{Init} initializes a choreography denoted by global type $\G$.  A network $\new(\G,\Lg)$ 
evolves in a reduction step into a composition of monitored processes and a session queue. For each $\pp \in \pt(\G)$,
there must be a pair $(\P_\pp, \T_\pp)$ in the collection
$\procset$. The type $\T_\pp$ must be adequate for the monitor obtained 
as projection of $\G$ onto $\pp$. Then the process (where the channel
$y$ has been replaced by $s[\pp]$) is coupled with
the
corresponding monitor and the empty queue $\ch : \eq$ is created.  The
security levels of the monitors are instantiated at runtime: while the
initial reading level is obtained from the mapping $\Lg$,
the initial writing level is $\bot$. Lastly, the name $\ch$ is
restricted. 
\item Rule \rln{UpLev} updates the current writing level $\wlev$ of the monitor
  with the least upper bound (join) of $\wlev$ and the
  level $\glev$ of the conditional expression tested by the process
  (by the semantics of processes, we know that $\glev$ is 
the level
  of a conditional expression). This is to prevent usual information leaks.
\item Rule \rln{In} defines the input of an extended value $\vo$. 
The input action must be enabled by the monitor. We further require 
 that the level associated with $\vo$ in the queue be lower than or
 equal to the reading level $\rlev$ of the monitor. 
 Notice that this check is nontrivial only for values, as nonces
have all level $\bot$.
\item 
  Rule \rln{Out} defines the output of an extended value $\vo$. If
  $\vo$ is a proper value $v$,
we require that the output be allowed by the monitor, i.e.,
  that the level associated with $v$ 
  be higher than or equal to the level $\wlev$ of the monitor. Nothing is
  required in case $\vo$ is a nonce, since nonces provide no information.

\item Rule \rln{InGlob} defines the \emph{global}
  reconfiguration mechanism for reading violations, which creates
    nonces.
A reading violation occurs when the level associated with the value
  in the queue is not lower than or equal to the reading level of the
  monitor. A reduction is still enabled, but since the monitored process is
  not allowed to input the provided value, an adaptation is realized
  by: (a) inputting a fresh nonce instead of the value, and
  (b) removing the unreadable value from the queue. 
In this rule and in the next one the function $\next(\Nonces)$ is used to obtain a fresh nonce in the set $\Nonces$.

\item Rule~\rln{OutGlob} defines the \emph{global}
  reconfiguration mechanism 
  for writing violations.  Such a violation occurs when the level of
  the sent value $\val$ (no writing violation may occur with nonces)
  is not greater than or equal to 
  the writing level $\wlev$ of the monitor controlling the sender
  (noted $\pp$ in the rule).
Also in this case a reduction is enabled; adaptation is realized by:
(a)~adding a fresh nonce to the queue and (b)~updating the reading
permission $\rlev$ attached to the monitor of $\pp$.  Indeed, to
formalize the fact that $\pp$ is responsible for the writing
violation, by trying to ``declassify'' value $\val$ from its
  original level to the reading level $\rlev'$ of the monitor
  controlling the receiver (noted $\q$ in the rule), we update its
current reading level $\rlev$ to the greatest lower bound (meet)
  of $\rlev$ and $\rlev'$. 
Hence, the reading level of $\pp$ is downgraded to that of $\q$ (or
lower), accounting for the fact that $\pp$ attempted to leak
information to $\q$. 
This is intended to counter any possible ``recidivism'' in
  $\pp$'s offending behaviour, by preventing 
new sensitive values to be received by $\pp$ and then leaked again to $\q$.

\item Rule \rln{Refresh} goes hand-in-hand with rules \rln{InGlob} and
  \rln{OutGlob}.
It extracts the set of participants whose processes
  can send $\nonce{i}$, which are the processes that contain
  $\nonce{i}$, and all those
which (transitively) communicate
  with them.
This set is obtained using the
  mapping $\A$.
  For the participants affected by $\nonce{i}$, a new global type is
  obtained via a function $F$.
  This function is left unspecified, for we are interested in
  modelling the mechanism of adaptation, and not the way in which the
  new security global type is chosen. Notice that the new
    security global type may involve other participants than those
    affected by $\nonce{i}$.  The reduction step then consists in (a)
    starting the new choreography and (b) continuing the execution of
    the unaffected participants. For (b) we must erase from the
    queue all messages involving affected participants; we denote by
   $\que \setminus \Pi'$ the resulting queue.

\item Rule \rln{InLoc} defines the \emph{local} reconfiguration
  mechanism triggered in the case of a reading violation.
  Intuitively, this rule defines adaptation by ``ignoring'' the
 forbidden input: the message is removed from the queue and the
  implementation of the monitored process is replaced with new code
  where the input action is not present.  This code replacement is
  formalized simply by considering the monitor that results from the
  reduction (noted $\widehat\M_\pp$ in the rule), and picking a process $P'$ that
  agrees with it.

\item Rule \rln{OutLoc} defines the \emph{local} reconfiguration
  mechanism for writing violations. 
As for \rln{InLoc},
 the monitor is modified and a new implementation that conforms to the
  modified monitor is injected. The monitor $\M_\q\setminus
  ?(\pp,\lambda)$ is obtained from $\M_\q$ by erasing the
  input action $?(\pp,\lambda)$ and choosing the corresponding branch.
The reading permission of the sender monitor is modified as in rule \rln{OutGlob}.
%
%

\item Rules \rln{Equiv} and \rln{Ctx} are standard: they allow the interplay of reduction
  with structural congruence and enable the reduction within
  evaluation contexts (defined as expected), respectively.
\end{enumerate}
The reduction of networks is clearly nondeterministic, in
contrast with standard session calculi. Nondeterminism arises
at every security violation, which can be treated either by generating
nonces (rules \rln{InGlob} and \rln{OutGlob}) or by
modifying the receiver's monitor and process,
just skipping
``wrong'' message receptions (rules \rln{InLoc} and \rln{OutLoc}). On
top of these alternatives, rule \rln{Refresh} can always be applied,
resulting in the splitting of the choreography between a part
affected by 
a fixed nonce (arbitrarily chosen) and 
an unaffected part.
As a result, the affected participants are
adapted using some (unspecified) adaptation function, while unaffected
participants remain unaware of this adaptation.

\bigskip

\noindent {\bf \emph{Main Results.}}
As in~\cite{CDV14}, well-typed networks
  enjoy \emph{subject reduction} and \emph{progress} properties. 
  Moreover, 
reduction of well-typed
networks always respects reading and writing permissions:



\begin{theorem}\label{main}
Let $\N$ be a network.
\begin{enumerate}[1.]
\item If $\N = \M_\pp^{\rlev,\wlev}[\P]\pc \ch:\mess\q\pp{\lambda(\val)}\cdot\que\red{} \widehat\M_\pp^{\rlev,\wlev}[\P']\pc \ch:\que$, then either $ lev(\val)\leq\rlev$ or $\P'$ is not obtained by consuming the message $\mess\q\pp{\lambda(\val)}$.
\item If $\N = \M_\pp^{\rlev,\wlev}[\P]\pc \ch:\que\red{} \widehat\M_\pp^{\rlev,\wlev}[\P']\pc \ch:\que\cdot\mess\pp\q{\lambda(\val)}$,  then $\wlev\leq lev(\val).$
\end{enumerate}
\end{theorem}
Theorem~\ref{main}(1) says that 
if the reading permission of a monitor is not respected,
then the disallowed value is never read from the queue---by virtue of the
runtime mechanisms implemented by rules \rln{InGlob} and
\rln{InLoc}. Analogously, Theorem~\ref{main}(2) says that if a value
is added to a session queue, then it is always the case that this is
allowed by the writing permission of the given monitor. Here again, it
is worth observing that adaptation mechanisms defined by rules
\rln{OutGlob} and \rln{OutLoc} can always be triggered to handle
the situations in which the sender attempts to transgress his
monitor's writing permission.


\section{Concluding Remarks}

Our work builds on~\cite{CDV14}, where a calculus based on global types, monitors and processes similar to ours 
was introduced. There are two main points of departure from that work. First, the calculus of~\cite{CDV14} relied on a global state, and  global types describe only finite protocols; adaptation was triggered after the execution of the communications prescribed by a global type, in reaction to changes of the global state. Second, adaptation in~\cite{CDV14} involved all participants in the choreography. In sharp contrast, in our calculus reconfigurations are triggered by security violations, and reconfiguration may be either local or global.
Therefore, we may consider our adaptation mechanism as more flexible than that of~\cite{CDV14} in two respects. First, adaptation is  triggered as a reaction to security violations (whose occurrence is hard to predict) rather than at fixed, prescribed computation points. Second, adaptation may be restricted to a subset of participants (those   involved in the security violation), thus resulting in a less disruptive procedure.

\bigskip

Our approach based on monitored processes (as defined in~\cite{CDV14}) relies on rather elementary assumptions on the nature of processes.
In particular, we assume that processes are well typed with respect to a rather simple discipline (based on intersection and union types) which does not mention security permissions. In fact, runtime information on permissions is handled by the monitor of the process; the relationship between typed processes and monitors is formalized by the notion of adequacy. This degree of independence between typed processes and security annotations distinguishes our approach from previous works on security issues for multiparty session typed processes (see, e.g.~\cite{DBLP:journals/corr/abs-1108-4465,CCD12}).

\paragraph{Acknowledgments.} 
We are grateful to the anonymous reviewers for their useful remarks.
This work was supported by COST Action IC1201: Behavioural Types for Reliable Large-Scale Software Systems 
via a Short-Term Scientific Mission grant (to P\'{e}rez). Dezani was also partially supported by MIUR PRIN Project CINA Prot. 2010LHT4KM and Torino University/Compagnia San Paolo Project SALT.

\bibliographystyle{eptcs}
\bibliography{adapt}

\end{document}